\documentclass[twocolumn,superscriptaddress,showpacs,prd,aps,amsmath,amssymb,nofootinbib]{revtex4-1}
\usepackage{natbib}
\usepackage{graphicx,color}
\usepackage{amsmath,amssymb}
\usepackage{verbatim}
\usepackage{float}
\usepackage{wasysym}
\usepackage{amssymb,graphicx}
\usepackage{epsfig}
\usepackage{psfrag}
\usepackage{dsfont}
\usepackage{amsfonts}
\usepackage{mathrsfs}
\usepackage{times}

\begin{document}

\title{General relativistic magnetohydrodynamics
  simulations of prompt-collapse neutron star mergers:
  The absence of jets}
\author{Milton Ruiz}
\affiliation{Department of Physics, University of Illinois at
  Urbana-Champaign, Urbana, IL 61801}
\author{Stuart L. Shapiro}
\affiliation{Department of Physics, University of Illinois at
  Urbana-Champaign, Urbana, IL 61801}
\affiliation{Department of Astronomy \& NCSA, University of
  Illinois at Urbana-Champaign, Urbana, IL 61801}


\begin{abstract}
  Inspiraling and merging binary neutron stars are not only
  important source of gravitational waves, but also promising
  candidates for coincident electromagnetic counterparts.
  These systems are thought to be progenitors of short
  gamma-ray bursts (sGRBs). We have shown previously that
  binary neutron star
  mergers that undergo {\it delayed} collapse to a black hole
  surrounded by a {\it weighty} magnetized accretion disk can
  drive magnetically-powered jets. We now perform
  magnetohydrodynamic simulations in full general relativity
  of binary neutron stars mergers that undergo {\it prompt}
  collapse to explore the possibility of jet formation from
  black hole-{\it light} accretion disk remnants. We find that
  after $t-t_{\rm BH}\sim 26(M_{\rm NS}/
  1.8M_\odot)$ms [$M_{\rm NS}$ is the ADM mass] following prompt
  black hole formation, there is no evidence of mass outflow
  or magnetic field collimation. The rapid formation of the
  black hole following merger prevents magnetic energy from
  approaching force-free values above the magnetic poles, which
  is required for the launching of a jet  by the usual
  Blandford--Znajek mechanism. Detection of gravitational waves
  in coincidence with sGRBs may provide constraints on the nuclear
  equation of state (EOS): the fate of an NSNS merger--delayed or
  prompt collapse, and hence the appearance or nonappearance of
  an sGRB--depends on a critical value of the total mass of the
  binary, and this value is sensitive to the EOS.
\end{abstract}

\maketitle

\section{Introduction}

The LIGO collaboration has reported the direct detection of
gravitational waves (GWs) from the inspiral and merger of at
least three binary black hole (BHBH) systems~\cite{Abbott:2016nhf,
  LIGO_first_direct_GW,Abbott:2016nmj,Abbott:2017}. Thus, it
may be just a matter of time before GWs from merging black
hole-neutron star (BHNS) and/or binary neutron stars (NSNS)
systems are detected as well. Estimates from population synthesis
and the current sensitive volume of the advance LIGO interferometers
predict detection rates  of $\lesssim 4$ events per year for
BHNS systems, and $\lesssim 20$  events per year for NSNS
systems~(see e.g. \cite{Aasi:2013wya,Abbott:2016ymx,
  Dominik:2014yma,aetal10}).

Merging BHNSs and NSNSs are not only important sources
of gravitational radiation, but also promising candidates
for coincident electromagnetic (EM) counterparts. These
systems have long been thought to be the progenitors of
{\em short} gamma-ray bursts (sGRBs)~\cite{Blinnikov84,EiLiPiSc,NaPaPi,
  Pacz86,Piran:2002kw,bergeretal05,Foxetal05,hjorthetal05,
  bloometal06,Baiotti:2016qnr,Paschalidis:2016agf}, which
is strongly supported by the {\it first} detection of a
kilonova associated with the sGRB ``GRB 130603B''
\cite{Tanvir:2013pia,Berger:2013wna}.

Coincident detection of GWs with EM signals from compact
binary mergers containing NSs could give new insight into
their sources: GWs are sensitive to the density profile
of NSs and their measurement enforces tight constraints
on the equation of state (EOS) of NSs~\cite{Lackey:2014fwa}.
Post-merger EM signatures, on the other hand, can help to
explain, for example, the phenomenology of sGRBs, and
the role of these BHNS and NSNS mergers
in triggering the nucleosynthesis processes in their ejecta
(e.g., the r-process; see  \cite{Sekiguchi:2015dma,
  Sekiguchi:2016bjd,Foucart:2016vxd}).

Recently, self-consistent simulations in full general
relativistic magnetohydrodynamics (GRMHD) of  merging
BHNSs~\cite{prs15} and merging NSNSs~\cite{Ruiz:2016rai}
that undergo {\it delayed} collapse have shown that when
the NSs are suitably magnetized, a collimated, mildly
relativistic outflow---an incipient jet can be launched
from the spinning BH remnant surrounded by a highly magnetized
accretion disk. In the BHNS scenario, the key ingredient
for jet launching is the existence of a strong poloidal
B-field component after disruption \cite{GRMHD_Jets_Req_Strong_Pol_fields,UIUC_PAPER2}.
This property can be achieved if initially the NS is
endowed with a dipole B-field that extends from the NS
interior into a pulsar-like exterior magnetosphere. Following
disruption, magnetic winding and the magnetorotational instability
(MRI) build up enough magnetic pressure above the BH poles
to allow the system to launch a jet  after~$\sim 100
(M_{\rm NS}/1.4M_\odot)$ms following the BHNS merger
\cite{prs15}. The burst duration and the outgoing Poynting
luminosity were found to be~$\Delta t\sim 0.5(M_{\rm NS}/
1.4M_\odot)$s and $L_{EM}\sim 10^{51}\rm erg\,s^{-1}$,
respectively, consistent with the observed duration of sGRBs
and their corresponding luminosities~\cite{Berger2014}
\footnote{See e.g.~https://swift.gsfc.nasa.gov/archive/grb$\_$table/fullview}.
In the NSNS scenario, by contrast, jets arise whether or not
the B-field is confined to the NS interior~\cite{Ruiz:2016rai}.
The key ingredient for jet launching seems to be  B-field
amplification due both to the Kelvin-Helmholtz instability
(KHI) and to the MRI, which can boost the rms value of the
B-field to~$\gtrsim 10^{15.5}$G~\cite{Kiuchi:2014hja,Kiuchi:2015sga}.
The calculations in~\cite{Ruiz:2016rai} showed that binary
NSNSs that start from the late inspiral  and undergo
{\it delayed} collapse to a BH launch jets after $\sim 44
(M_{\rm NS}/1.8M_\odot)$ms following the NSNS merger. The burst
duration and its EM luminosity were found to be $\Delta t\sim
97(M_{\rm NS}/1.8M_\odot)$ms and $L_{EM}\sim 10^{51}\rm erg\,
s^{-1}$, respectively, also consistent with short sGRBs; see
e.g.~\cite{Kann:2008zg}.

Although the above results were obtained using a simple, $\Gamma$-law
EOS, it is expected that a realistic EOS will yield a similar outcome.
Different EOSs affect the amount and composition of the ejecta during
NSNS coalescences~\cite{Hotokezaka2013,Palenzuela:2015dqa,Ciolfi:2017uak,
  Sekiguchi2015,Foucart:2016rxm,Lehner:2016lxy}, and  therefore, the
ram pressure produced by the fall-back debris, as well as the mass
of the accretion disk. The delay time for jet launching following the
merger may therefore depend on the EOS. Moreover different EOSs have
different threshold masses above which the collapse is prompt vs.
delayed~\cite{Shibata:2002jb,STU1,ST}. However, for all EOSs the most
significant feature that determines whether jets can be launched is likely
whether the merger remnant undergoes delayed or prompt collapse,
although even in the delayed collapse, different EOSs have strong
impact on the accretion disk~\cite{Shibata:2006nm,Rezzolla:2010fd},
and hence in the jet's lifetime.
The above result seems to be the
main reason why in the much higher resolution but shorter NSNS
simulations reported in~\cite{Kiuchi:2014hja}, in which a
H4-EOS is assumed and the B-field is confined to the NS interior,
neither a magnetically driven outflow or a B-field collimation were
observed. After $t\sim 26$ms following the BH formation, fall-back
material in the atmosphere persisted. It is likely then that,
at that point in the evolution, the ram pressure is still
larger than the magnetic pressure and thus a longer simulation
is  required for the jet to emerge.
While in the NSNS simulations reported
in~\cite{Kawamura:2016nmk}, in which the effects of different EOSs,
different mass ratios, and different B-field orientations were studied,
there is no evidence of an outflow or a jet, there is a  formation
of an organized B-field structure above the BH. Therefore, we expect
that, in a longer simulation,  a jet may be launched. See
also~\cite{Ciolfi:2017uak} for a detailed discussion of
the rotation profiles, the accretion disk evolution and
amplification of the B-field, as well as the ejection of matter
in magnetized merging NSNSs. Note also that neutrino
pair annihilation alone may not be strong enough to power jets
\cite{Just:2015dba,Perego:2017fho}.

To complete our preliminary survey of NSNS mergers as possible
sGRB progenitors, we now
consider magnetized NSNS configurations that lead to {\it prompt}
collapse following merger. These events produce less massive
accretion disks than those arising from delayed collapse~(see
e.g.~\cite{STU1,Shibata:2006nm,Rezzolla:2010fd}). For comparison
purposes, we again consider NSNS binaries  described initially by
irrotational $\Gamma=2$ polytropes endowed with the same
two  B-field configurations employed in \cite{Ruiz:2016rai}.

We find, in agreement with previous studies~\cite{STU1,Shibata:2006nm,
  Rezzolla:2010fd,Liu:2008xy,Duez2010CQGra}, that prompt collapse
leads to a highly spinning BH remnant ($a/M_{BH} \gtrsim 0.8$),
with an accretion disk mass much smaller than $0.01M_\odot(k/262.7\rm
km^2)^{1/2}$, and increasing with greater disparity between the rest
masses of the two NSs.  Here $k$ is the polytropic gas constant:
$k=P/\rho_0^\Gamma$. Thus, these
results are not altered by the presence of weak interior B-fields.
We now also find that, in contrast to delayed collapse, the absence
of a hypermassive neutron star (HMNS) epoch  does not allow the
magnetic energy to reach equipartition and, ultimately, force-free
levels~\cite{Kiuchi:2015sga}, thereby preventing B-field collimation
along the remnant BH poles and an associated jet outflow.

Although our study is far from exhaustive, we tentatively conclude
that GWs from merging NSNSs may be accompanied by sGRBs in the case
of delayed collapse but not in the case of prompt collapse. This
finding has important consequences. The fate of a NSNS merger
-- prompt  or delayed collapse -- is determined by a critical
value of the total mass, and this value depends on the EOS~\cite{ST,
  Shibata:2005ss}.
If the masses of the NSs in the binary can
be reliably determined from measurements of the GWs during the
pre-merger inspiral phase~\cite{Cutler:1994ys,
  Sathyaprakash:2009xs}, then the absence or presence of a
counterpart sGRB following merger will shed light on the EOS. This
information may supplement other estimates
of stellar radii and compactions from tidal imprints in the waveforms
\cite{Maselli:2013rza,Hinderer:2009ca}. Additionally, measurement
of the time delay between the peak GW  and sGRB signals may help provide
an estimate of the initial NS B-field strength~\cite{BaShSh}.

The paper is organized as follows. A short summary of the numerical
methods and their implementation is presented in Sec.
\ref{subsec:nume_setup}. A detailed description of the adopted initial
data and the grid structure  used for solving the GRMHD equations
are given in Sec.~\ref{subsec:idata} and Sec.~\ref{subsec:grid},
respectively. Sec.~\ref{subsec:diagnostics} contains the diagnostics
employed to monitor our numerical calculations. We present our results
in Sec.~\ref{sec:results}. Finally, we  offer conclusions in Sec.
\ref{sec:conclusion}. We adopt geometrized units ($G=c=1$) throughout
the paper, unless otherwise specified.

\section{Numerical Methods}
\label{sec:Methods}

\subsection{Numerical setup}
\label{subsec:nume_setup}
We use the Illinois GRMHD code, which is embedded in the
\texttt{Cactus}\footnote{http://www.cactuscode.org}
infrastructure and uses \texttt{Carpet}\footnote{
  http://www.carpetcode.org} for moving mesh refinement.
This code has been thoroughly tested and used in the past
in numerous scenarios involving compact objects, including
magnetized BHNS and NSNS simulations~(see e.g.
\cite{Etienne:2007jg,Etienne:2011ea,prs15,Ruiz:2016rai}).
A detailed description of the numerical methods, their
implementation, and code tests can be found in, e.g.,
\cite{Etienne:2011ea,Etienne:2011re,Etienne:2010ui,
  Etienne:2011re}.

\begin{figure}
  \centering
  \includegraphics[width=0.45\textwidth]{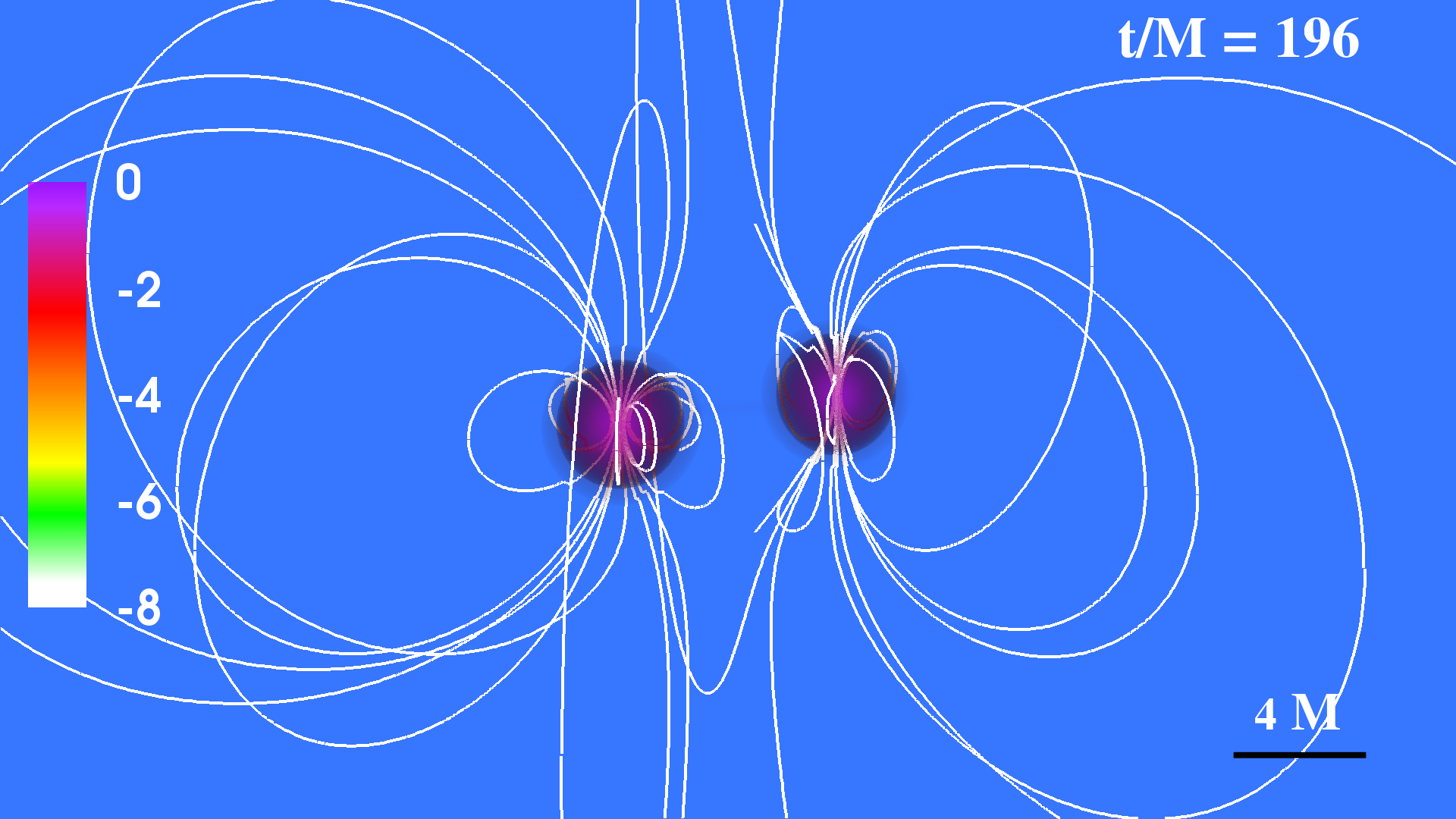}
  \includegraphics[width=0.45\textwidth]{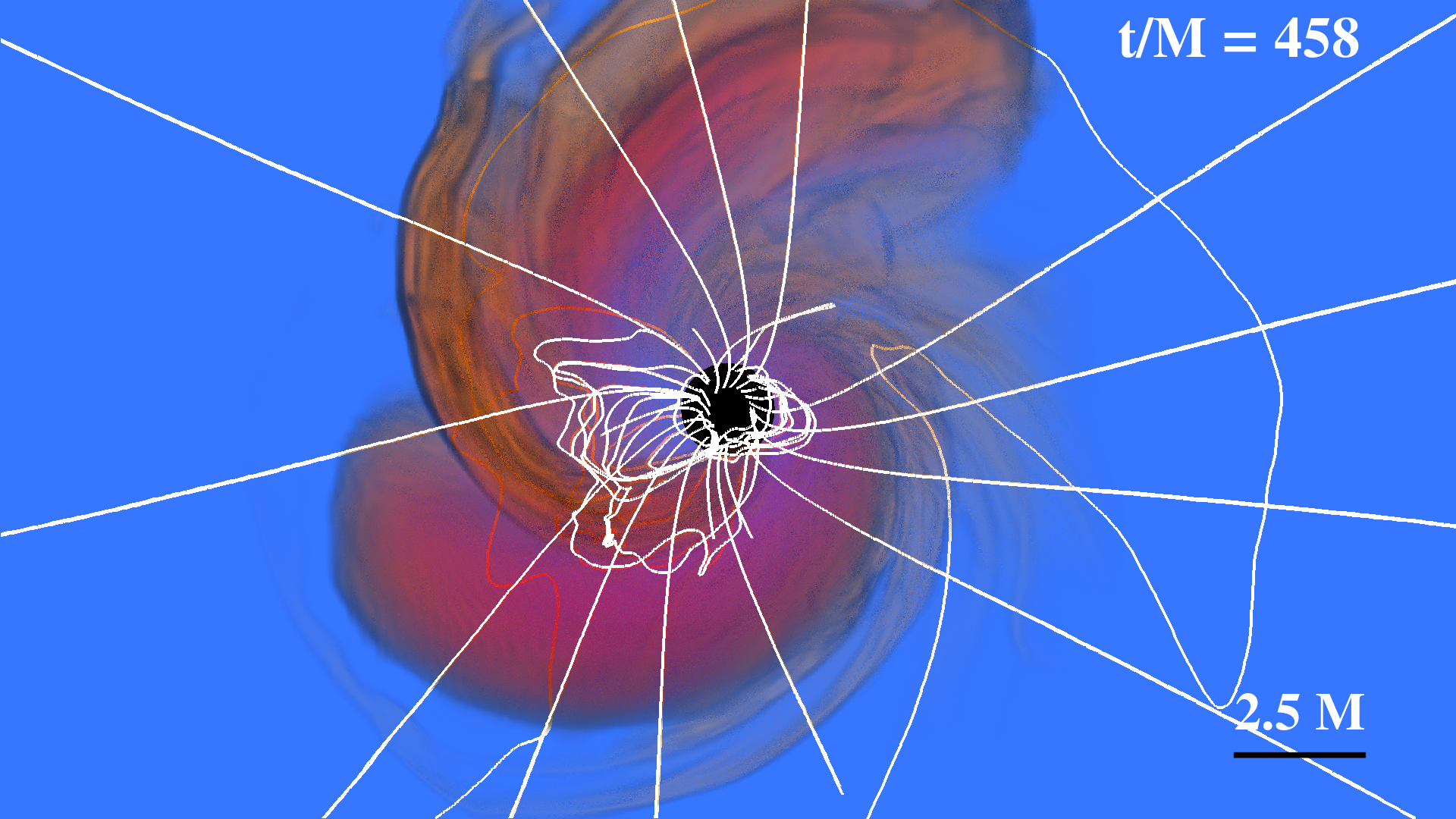}
  \includegraphics[width=0.45\textwidth]{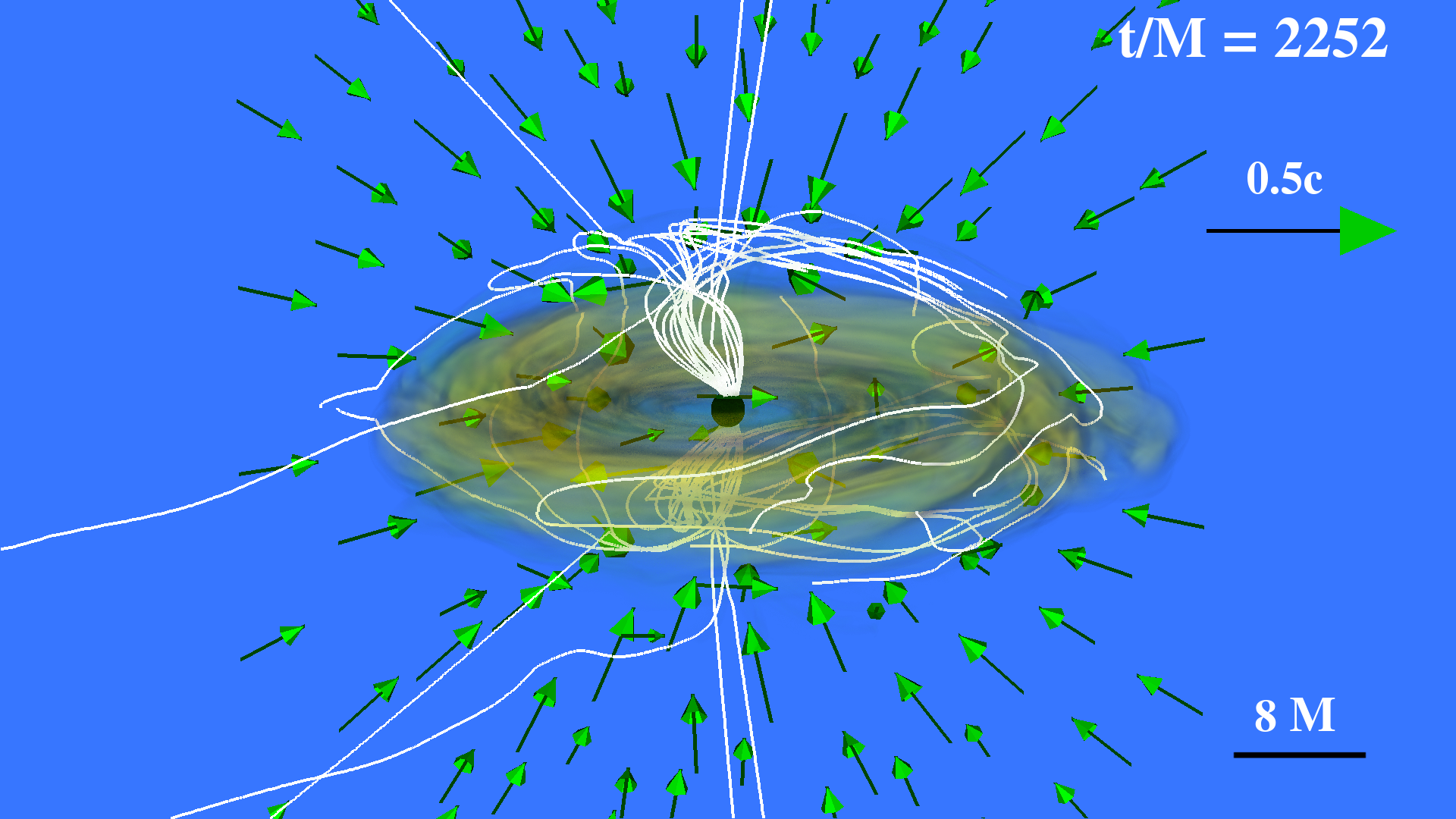}
  \caption{Volume rendering of rest-mass density $\rho_0$ normalized to its initial
    maximum value  $\rho_{0,max}=7.9\times 10^{14}(1.8M_\odot/M_{NS})^2\,{\rm g/cm}^3$
    (log scale) at selected times for the P-Prompt-3 case (see Table~\ref{table:ID_allcases}).
    The top panel shows the time at which the stars are seeded with the  B-field (white lines)
    generated by the vector potential $A_\phi$ in Eq.~(\ref{eq:Vector_potetail}), the  middle
    panel shows a top view during the BH formation (the black sphere), and the bottom panel
    shows the  end of the simulation. Arrows indicate plasma velocities.  Here  $M=1.1\times
    10^{-2}(M_{\rm NS}/1.8M_\odot)\rm ms$ = $3.31(M_{\rm NS}/1.8M_\odot)\rm km$.
    \label{fig:NSNS_ID}}
\end{figure}

\paragraph*{\bf Spacetime evolution:}
We decompose the metric into  $3+1$ form,
\begin{align}
  ds^2&=-\alpha^2\,dt^2+\gamma_{ij}\,\left(dx^i+\beta^i\,dt\right)
  \left(dx^j+\beta^j\,dt\right)\,,
\end{align}
with $\alpha$ and  $\beta^i$ the familiar gauge variables
and $\gamma_{ij}$ the spatial metric induced on a spatial
hypersurface
with a future-directed, timelike unit vector $n^\mu=
(1/\alpha,-\beta^i/\alpha)$.  The full spacetime metric
$g_{\mu\nu}$ is related to the spatial metric by $\gamma_{\mu\nu}=
g_{\mu\nu} +n_\mu\,n_\nu$.
Associated with the time slice, we define  the extrinsic
curvature $K_{\mu\nu}\equiv-\gamma_{\mu\alpha}\nabla^\alpha
n_\nu$. Geometric  variables are then  evolved via the
Baumgarte--Shapiro--Shibata--Nakamura (BSSN) formulation
\cite{shibnak95,BS}. The resulting evolved
variables are then the  conformal exponent $\phi=\rm{ln}(\gamma)/12$,
conformal metric $\tilde{\gamma}_{ij}=e^{-4\phi}\gamma_{ij}$, the
trace of the extrinsic curvature $K$, the conformal trace-free
extrinsic curvature $\tilde{A}_{ij}=e^{-4\phi}(K_{ij}-\gamma_{ij}\,
K/3)$ and the three auxiliary variables $\tilde{\Gamma}^i=-\partial_j
\tilde{\gamma}^{ij}$. We evolve these variables using the
equations of motion (9)-(13) in~\cite{Etienne:2007jg}. We close
the system of equations of motion in the geometric sector by using
$1+$log time slicing and the gamma-driver spatial shift conditions
\cite{Bona:1994dr,Alcubierre:2002kk} cast in first order
form~(see e.g.~\cite{Ruiz:2010qj}).
For numerical stability, we set the damping parameter $\eta$
appearing in the shift condition to $\eta=0.85/M$, with $M$ the
ADM mass of the system.

The spatial discretization is performed by using  fourth-order
accurate, cell-centered, finite-differencing stencils, except
on shift advection terms, where fourth-order accurate upwind
stencils are used~\cite{Etienne:2007jg}. Outgoing wave-like
boundary conditions are applied to all the evolved variables.
The time integration is performed via the method of lines using
a fourth-order accurate, Runge-Kutta integration scheme.

%
\begin{center}
  \begin{table*}[th]
    \caption{Initial data for the NSNS prompt collapse cases, as well as
      the delayed collapse case considered in~\cite{Ruiz:2016rai}. All the models
      have an initial separation of $44.42\,(k/k_0)^{1/2}$km, where $k_0=262.7\rm km^2$.
      Columns show the compaction $(M/R)_i$ of each companion $i=1,2$, which is computed
      assuming an isolated spherical star with the same rest mass, the coordinate equatorial
      radius of each star $R_{{\rm eq}_i}$, the total rest mass $M_0$, the  ADM mass $M_{\rm ADM}$,
      the  ADM angular momentum $J_{\rm ADM}$,  and the binary angular frequency
      $\Omega$. These models are also listed in Tables~III-IV of~\cite{tg02}. For completeness,
      we include the initial magnetic energy in units of $10^{50}\rm erg\,s^{-1}$ as defined
      in~Eq.~(\ref{eq:EM_energy}) for models P and I, respectively.
      \label{table:ID_allcases}}
    \begin{tabular}{cccccccccc}
      \hline\hline
          {Model} & $(M/R)_1$ &$(M/R)_2$ &$R_{\rm eq_1}(k/k_0)^{1/2}$ &$R_{\rm eq_2}(k/k_0)^{1/2}$ &$M_0$ $(k/k_0)^{1/2}$& $M_{\rm ADM}$ $(k/k_0)^{1/2}$& $J_{\rm ADM}$ $\,(k/k_0)$ &  $\Omega$ $(k_0/k)^{1/2}$ &
          $\mathcal{M}$ \\
          \hline
          Prompt-1  &  0.16   &0.16  &12.2 &12.2 km &3.51$M_\odot$  & 3.22$M_\odot$  & 9.87$M_\odot^2$   & 1914.7 $\rm{s}^{-1}$ & $1.2,\,1.4$\\
          Prompt-2  &  0.18   &0.18  &11.0 &11.0 km &3.75$M_\odot$  & 3.40$M_\odot$  & 10.90$M_\odot^2$  &2218.6 $\rm{s}^{-1}$  & $2.0,\,1.8$\\
          Prompt-3  &  0.16   &0.18  &12.2 &11.1 km &3.63$M_\odot$  & 3.31$M_\odot$  & 10.37$M_\odot^2$  &2188.2 $\rm{s}^{-1}$  & $1.7,\,1.8$\\
          Delayed   &  0.14   &0.14  &13.5 &13.5 km &3.20$M_\odot$  & 2.96$M_\odot$  & 8.61$M_\odot^2$   & 1884.3 $\rm{s}^{-1}$ & $1.4,\,3.2$\\
          \hline\hline
    \end{tabular}
  \end{table*}
\end{center}

\paragraph*{\bf MHD evolution:}
The Illinois code solves the equations of ideal GRMHD in a
conservative scheme via high-resolution shock capturing
methods to handle shocks~\cite{Duez:2005sf}. For that it
adopts the conservative variables
\cite{Etienne:2010ui}
\begin{align}
  \rho_* \equiv &- \sqrt{\gamma}\, \rho_0\,n_{\mu}\,
  u^{\mu}\,,\\
  \tilde{S}_i \equiv&-\sqrt{\gamma}\, T_{\mu \nu}\,
  n^{\mu}\,\gamma^{\nu}_i\,,\\
  \tilde{\tau}\equiv&\sqrt{\gamma}\, T_{\mu \nu}\,
  n^{\mu}\,n^{\nu} - \rho_*\,,
  \label{eq:conserv_varaibles}
\end{align}
where
\begin{align}
  T_{\mu \nu} = (\rho_0\,h+b^2)\,u_\mu\,u_\nu +
  \left( P + \frac{b^2}{2}
  \right)\,g_{\mu \nu} - b_{\mu}\,b_{\nu}\,,
  \label{eq:Mag_Tmunu}
\end{align}
is the stress-energy tensor for a magnetized plasma
with rest-mass density $\rho_0$, pressure $P$, specific
enthalpy~$h=1+\epsilon+ P/\rho_0$,  specific internal
energy $\epsilon$, B-field $b^\mu=B^\mu_{(u)}/(4\,
\pi)^{1/2}$ as measured by an observer co-moving with
the fluid, and $u^\mu$ the fluid four-velocity. The
resulting equations of motion are obtained via the
rest-mass and energy-momentum conservation laws
(see Eqs. (27)-(29) in~\cite{Etienne:2010ui}). To
guarantee that the B-field remains divergenceless
during the whole evolution, the code solves the
magnetic induction equation using a vector potential
$\mathcal{A}^\mu$~(see Eqs. (8)-(9) in
\cite{Etienne:2011re}). We also adopt the generalized
Lorenz gauge~\cite{Farris:2012ux,Etienne:2011re}
with a damping parameter~$\xi=16/M$. This gauge
avoids the development of spurious B-fields that arise
due to interpolations across the refinement levels in
moving-box simulations.
As pointed out in
\cite{Giacomazzo:2012iv,Etienne:2011re}, interpolations
at moving-box boundaries in $\mathcal{A}^\mu$-evolution
codes may produce spurious conversion of EM gauge
modes into physical modes and vice-versa, and as a
result spurious B-fields will eventually contaminate
the evolution~\cite{Etienne:2011re}. We close the
system of equations in the MHD sector by using a
$\Gamma-$law equation of state $P=(\Gamma-1)\rho_0\,
\epsilon$, with $\Gamma=2$ to model the NS matter.
Finally,  as is done in many hydrodynamic and ideal MHD
codes, we add a tenuous atmosphere $\rho_{\rm atm}$ in
the grid points where the rest-mass density is below a
threshold value. We set $\rho_{atm}=10^{-10}\,\rho_{max}$,
where $\rho_{max}$ is the maximum value of the initial rest-mass
density of the system~\cite{Etienne:2010ui}.

\subsection{Initial data}
\label{subsec:idata}
NSNS mergers may yield a remnant that can either form a transient
differentially-rotating HMNS that can survive for many rotation
periods~\cite{BaShSh}, or promptly collapse to a BH. The above
outcome depends strongly of the {\it total} mass of the system,
and independently of the mass ratio. If the total rest mass of a
$\Gamma=2$ EOS NSNS binary is $\gtrsim 3.44M_\odot(k/262.7 \rm
km^2)^{1/2}$, then  the system will  promptly collapse to a BH.
This mass corresponds to $\sim 1.7$ times the maximum allowed
rest mass of a single spherical NS, which turns out to be
$M_{\rm sph}\approx 1.98M_\odot(k/262.7 \rm km^2)^{1/2}$, or a
total ADM mass of $M^{\rm ADM}_{\rm sph}=1.8 M_\odot(k/262.7
\rm km^2)^{1/2}$.  Note that these results scale with
the polytropic constant $k=P/\rho_0^\Gamma$, which determines
the Tolman–Oppenheimer–Volkoff (TOV) maximum mass. Besides, the
threshold mass depends strongly
on the EOS. Namely, for  realistic EOSs, such as APR or
SLy, the threshold mass is $\sim1.3-1.35\,M_{\rm sph}$ (see e.g.
\cite{Shibata:2002jb,STU1,ST}).

We consider NSNS binaries in quasiequilibrium circular
orbits that inspiral, merge and undergo prompt collapse.
The initial stars are irrotational, $\Gamma=2$ polytropes,
and we evolve the matter with a $\Gamma$-law EoS, allowing
for  shock heating.
The initial data are computed using the publicly available
{\tt LORENE} code~\footnote{
  http://www.lorene.obspm.fr}.  All our models have an initial separation
of $44.42\,(k/262.7 \rm km^2)^{1/2}$km. Table~\ref{table:ID_allcases}
summarizes the initial parameters of the models considered. For
comparison purposes, we also include the NSNS delayed case treated
previously in~\cite{Ruiz:2016rai}.

\begin{figure}
  \centering
  \includegraphics[width=0.48\textwidth]{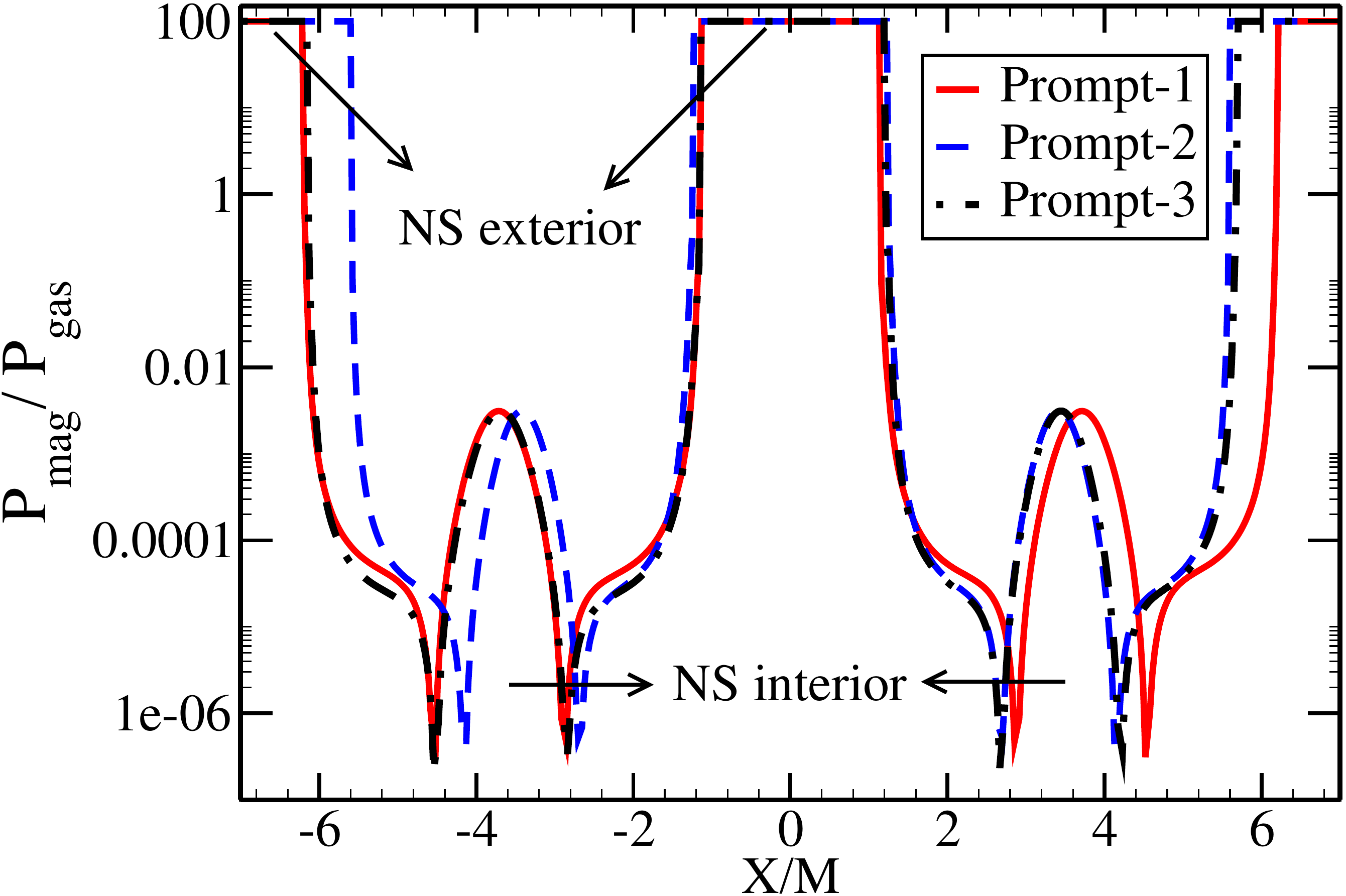}
  \caption{Magnetic-to-gas pressure ratio $\beta^{-1}\equiv P_{\rm mag}/
    P_{\rm gas}$ at the time a dipole-like B-field is seeded in the
    NSNS cases listed in Table~\ref{table:ID_allcases}. The B-field
    is generated by the vector potential~$A_\phi$ given by Eq.
    (\ref{eq:Vector_potetail}). Here the two NS are centered at $x/M=\pm\,
    3.45$ in the Prompt-1 case, and at $x/M=\pm\,3.67$ in the Prompt-2
    case, and $y/M=z/M=0$.  Notice that the position of each of the NSs
    in the Prompt-3 case matches the position of the corresponding
    companion with the same mass and compaction as in the above cases
    (see Table~\ref{table:ID_allcases}).
    \label{fig:plasma_init}}
\end{figure}

As in~\cite{prs15,Ruiz:2016rai}, and to avoid build up of numerical
errors, we evolve the above initial data until approximately two
orbits before merger. At that time, $t=t_B$, the NSs are endowed
with a dynamically unimportant interior B-field using one of the following
two prescriptions:

\begin{itemize}
\item {\bf Pulsar case}: In the pulsar case (hereafter the P case),
  the stars are seeded  with a dipolar B-field generated by the vector
  potential~\cite{Farris:2012ux,Paschalidis:2013jsa}
  \begin{align}
    A_\phi&= \frac{\pi\,\varpi^2\,I_0\,r_0^2}{(r_0^2+r^2)^{3/2}}
    \left[1+\frac{15\,
        r_0^2\,(r_0^2+\varpi^2)}{8\,(r_0^2+r^2)^2}\right]\,,
    \label{eq:Vector_potetail}
  \end{align}
  that approximately corresponds to that generated by an interior
  current loop (see top panel of Fig.~\ref{fig:NSNS_ID}).  Here $r_0$
  is the current loop radius, $I_0$ is the current,  $r^2=\varpi^2+z^2$,
  with  $\varpi^2= (x-x_{\rm NS})^2+(y-y_{\rm NS})^2$, and $(x_{\rm NS},
  y_{\rm NS})$ is the position of the NS centroid. We choose the current
  $I_0$ and radius of the loop $r_0$ such that the maximum value of the
  magnetic-to-gas-pressure ratio in the NS interior is $\beta^{-1}\equiv
  P_{\rm mag}/P_{\rm gas}=0.003125$ (see Fig.~\ref{fig:plasma_init})
  which matches the value used in~\cite{Ruiz:2016rai}.
  The resulting
  B-field strength at the NS pole turns out to be ${B}_{\rm pole}\approx
  1.58\times 10^{15}(1.8M_\odot/M_{\rm NS})$G. This B-field was chosen in
  \cite{Ruiz:2016rai} so that the rms value of the B-field in the transient
  HMNS is similar to that reported in the very high resolution NSNS
  simulations~\citep{Kiuchi:2015sga}, where it was shown that during the
  NSNS merger both the KHI and the MRI can boost the B-field strength to
  values from $\sim 10^{13}$G to $\sim 10^{15.5}$G, with local values
  up to $\sim 10^{17}$G.
  Finally, to reliably evolve the B-field in the stellar exterior and,
  at the same time,  mimic the low $\beta_{\rm ext}$ environment  that
  characterizes a force-free, pulsar-like magnetosphere, a new tenuous
  and variable density atmosphere satisfying $\beta^{-1}_{\rm ext}=100$
  everywhere in the exterior is initially imposed at $t=t_B$ on top of
  $\rho_{atm}$, as we did in~\cite{prs15,Ruiz:2016rai} (see Fig.~\ref{fig:plasma_init}).
  Since the dipole B-field strength falls away from the NS surface as $1/r^3$,
  this prescription forces the variable density in the atmosphere also to
  fall initially  as $1/r^3$ until it equals $\rho_{atm}$. Subsequently,
  all the dynamical variables, interior and exterior, are evolved according
  to the ideal GRMHD equations.
  This artificial atmosphere increases the total
  rest-mass of the system by less than~$\sim 1.0\%$.
  %
  \begin{center}
    \begin{table}[th]
      \caption{List  of  grid  parameters  for  all  models listed
        in Table~\ref{table:ID_allcases}.  The  computational mesh
        consists of two sets of seven nested grids centered one on
        each of the  NSs.
        Here $\Delta x_{\rm max}$ is the coarsest grid spacing.
        The grid spacing of all other levels is  $\Delta x_{\rm max}/2^{n-1}$
        with $n=2,\cdots,7$. Here $M$ is the total rest-mass of the system.
        \label{table:grid_allcases}}
      \begin{tabular}{ccccc}
        \hline\hline
            {Model}    & &$\Delta x_{\rm max}$ & &Grid hierarchy\\
            \hline
            Prompt-1   & & $1.84M$  & &$213.22M/2^{n-1}$\\ 
            Prompt-2   & & $1.56M$  & &$182.35M/2^{n-1}$\\ 
            Prompt-3   & & $1.78M$  & &$207.04M/2^{n-1}$\\ 
            Delayed    & & $2.20M$  & &$245.66M/2^{n-1}$\\ 
            \hline\hline
      \end{tabular}
    \end{table}
  \end{center}
  %
\item {\bf Interior case}: In the interior case (hereafter the I case) the
  stars are seeded with a poloidal B-field
  confined to the NS interior through the vector
  potential~\cite{Etienne:2011ea}
  \begin{align}
    A_i&= {C_i}\,\varpi^2\,{\rm max}(P-P_{\rm cut},0)^{n_b}~\,,
    \label{eq:Vector_potetail_int}
  \end{align}
  with
  \begin{align}
    C_i&=\left(
    -\frac{y-y_{NS}}{\varpi^2}\,{\delta^x}_i
    +\frac{x-x_{NS}}{\varpi^2}\,{\delta^y}_i
    \right)\,A_b\,,
  \end{align}
  where $A_b$, $n_b$ and $P_{\rm cut}$ are free parameters that
  parametrize the strength, the degree of central  condensation and
  the confinement of the B-field,  respectively. We set
  $P_\mathrm{cut} =0.01\,{\rm max}(P)$ and  $n_b =1$, and then $A_b$
  is chosen so that the resulting B-field at the center of each star
  coincides with that in the P-cases.
\end{itemize}

In all our cases (see Table~\ref{table:ID_allcases}), the
 magnetic-to-orbital-binding-energy ratio as defined in~\cite{Bejger:2004zx}
 is $\sim 1.45\times 10^{-6}$ and the  magnetic
dipole moment is aligned with the orbital angular momentum of the
system.

%
\subsection{Grid structure}
\label{subsec:grid}
The numerical grids consist of two sets of refinement boxes centered
on each of the NSs. Once they overlap they are replaced by a common
box centered on the system center of mass. Each set consists of seven
boxes that differ in size and in resolution by factors of two. The
finest box around each NS has a side half-length of $\sim 1.3\,
R_{\rm NS}$, where $R_{\rm NS}$ is the initial NS equatorial radius
(see Table~\ref{table:ID_allcases}). The  grid structure of the mesh
refinement used in the simulations is listed in~Table~\ref{table:grid_allcases}.
In all cases, the initial NS diameter is resolved  by   $\sim 180$ grid
points. We impose reflection symmetry across the orbital plane. Note
that this resolution matches the high resolution used in
\cite{Ruiz:2016rai}.
%
\subsection{Diagnostic quantities}
\label{subsec:diagnostics}
%
During the whole evolution we monitor several diagnostic quantities
to verify the reliability of our numerical calculations. We monitor
the normalized Hamiltonian and momentum constraints  defined
by Eqs.~(40)-(43) in~\cite{Etienne:2007jg}. In all cases listed in
Table~\ref{table:ID_allcases}, the constraint violations are below
$0.03$ throughout the evolution. As expected the constraints
peak during  BH formation and then decrease as the evolution
proceeds. We use the {\tt AHFinderDirect} thorn~\cite{ahfinderdirect}
to locate and monitor the apparent horizon. To estimate the BH mass
$M_{\text{BH}}$ and its dimensionless spin parameter $\text{a}_{BH}/
M_{\text{BH}}$ we use Eqs.~(5.2)-(5.3) in~\cite{Alcubierre:2004hr}.

To measure  the energy and angular momentum radiated in
form of GWs, we use a  modified version of the {\tt Psikadelia}
thorn that computes the  Weyl scalar $\Psi_4$, which is decomposed
into $s=-2$ spin-weighted spherical harmonics (see e.g.
\cite{Ruiz:2007yx}) at different radii between $\approx 30M\sim
135 (M_{\rm NS}/1.8M_\odot)$km and $\approx 160M\sim 710
(M_{\rm NS}/1.8M_\odot)$km. We find that between
$\sim 0.2\%$ and $\sim0.4\%$ of the energy of our models is
radiated away in form of gravitational radiation, while
between $\sim 1.8\%$ and $\sim 3.4\%$ of the angular momentum is
radiated. We also compute the outgoing EM (Poynting) luminosity
$L_{EM}\equiv-\int T^{r(EM)}_t\sqrt{-g}\,d\mathcal{S}$ across a given
surface $\mathcal{S}$, where $T^{(EM)}_{\mu\nu}$ is the electromagnetic
energy-momentum tensor. Using Eqs. (21)-(22) in~\cite{Etienne:2011ea},
and taking into account the GW and EM radiation losses, we verify the
conservation of the total mass $M_{\rm int}$ and the total angular
momentum $J_{\rm int}$, which coincide with the ADM mass and ADM
angular momentum of the system at spatial infinity.  In all cases the
total mass is conserved to within $\sim 1\%$, and  the total angular
momentum is conserved to within $\sim 8\%$.  Finally, we monitor the
magnetic energy
\begin{equation}
  \mathcal{M}=\int u^\mu u^\nu T^{(EM)}_{\mu\nu}\,dV\,,
  \label{eq:EM_energy}
\end{equation}
measured by a comoving observer~\cite{Etienne:2011ea}. Here $dV=e^{6\phi}
\,d^3x$ is the proper volume element on a spatial slice. Once the BH forms,
$\mathcal{M}$ is calculated in the fluid exterior of the horizon.
%
\begin{figure}
  \centering
  \includegraphics[width=0.49\textwidth]{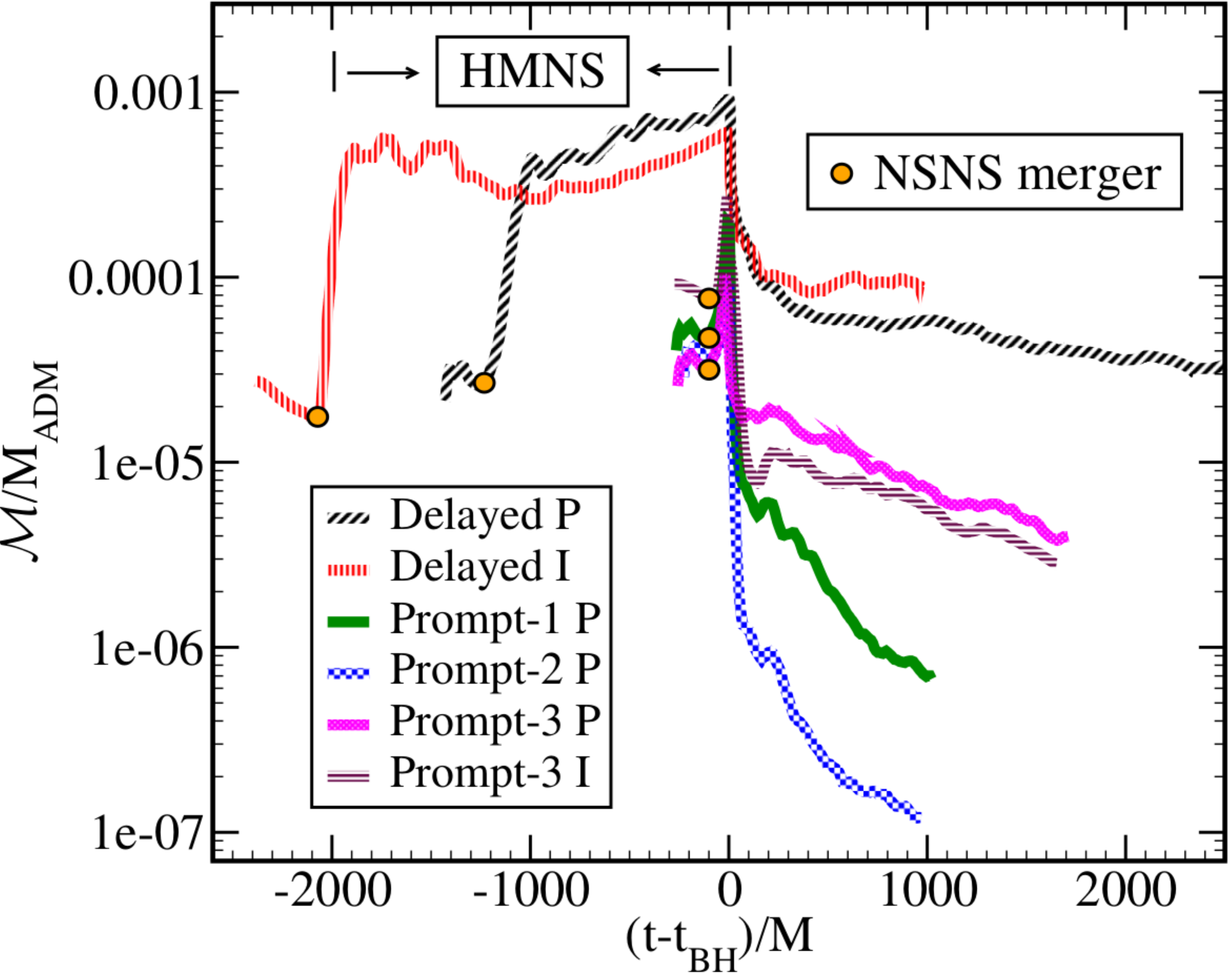}
  \caption{Total magnetic energy $\mathcal{M}$ (normalized by the
    ADM mass) vs. time for  cases in Table~\ref{table:ID_allcases}.
    Dots indicate the NSNS merger time. The horizontal axis has been
    shifted to the BH formation time. In contrast to the delayed collapse,
    magnetic instabilities during the HMNS epoch steeply amplify the magnetic
    energy $\mathcal{M}$. 
    \label{fig:NSNS_EM}}
\end{figure}
%
\begin{figure*}
  \centering
  \includegraphics[width=1.07\textwidth,trim=0 0 0 0,clip=true]{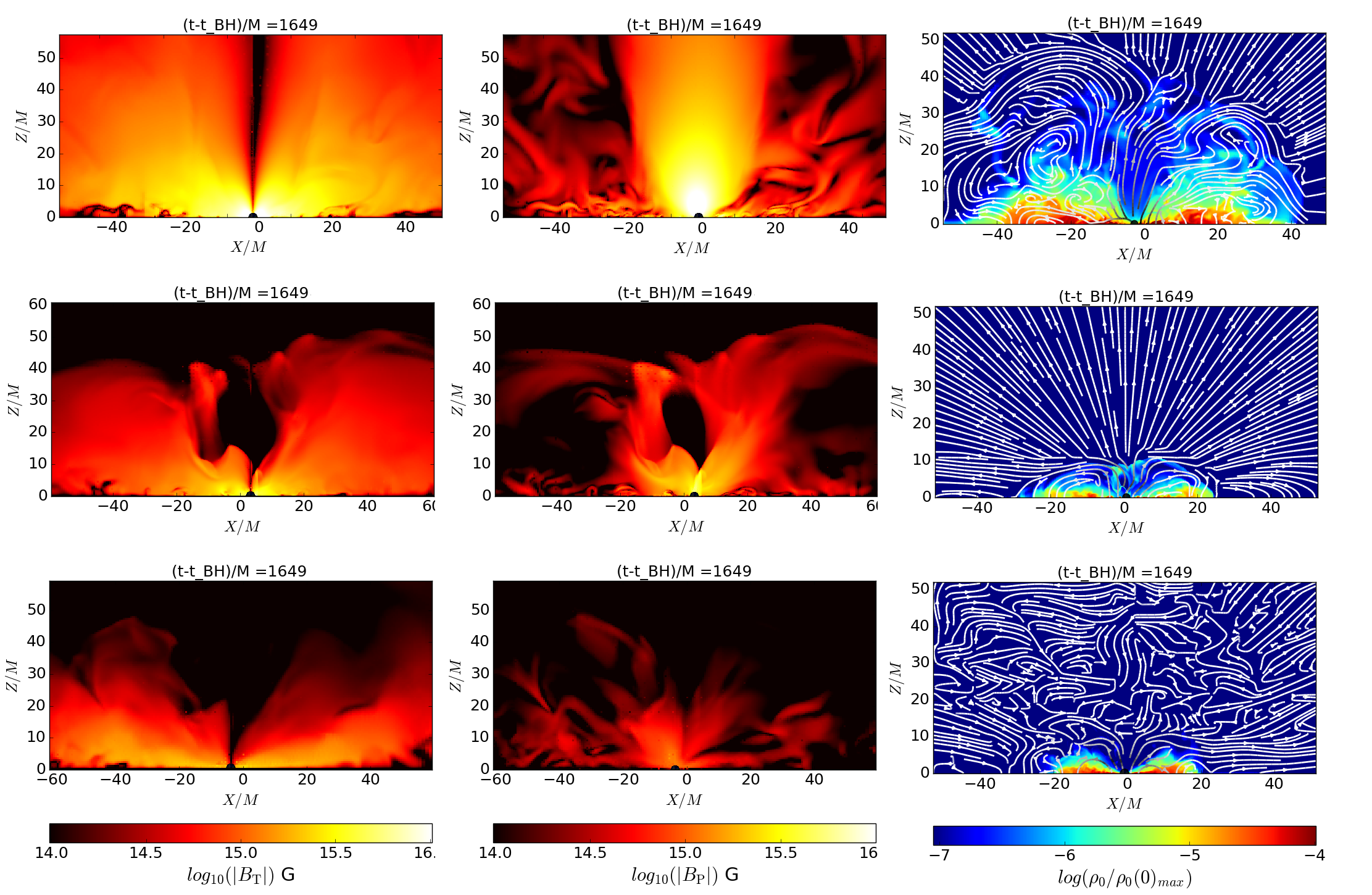}
  \caption{
    \label{fig:Tor_polBfield}
    Toroidal (left column) and poloidal (middle column) components of the B-field, and
    rest-mass density normalized to its maximum value (right column) on a meridional
    plane near the end of the prompt-3 cases. Top panel shows the P-Delayed  case,
    while the P-Prompt-3 and the I-Prompt-3 cases are shown in the middle and the bottom
    panels, respectively. The white lines on the right panel show the B-field structure,
    while the central black disks in all panels denote the apparent horizons. Notice that
    only in the delayed collapse case, the two components of the B-field have a strength
    $\gtrsim 10^{15.7}\,(1.8M_\odot/M_{NS})$G.}
\end{figure*}
%
\section{Results}
\label{sec:results}
The evolution of our binary NS models can be characterized by three
phases: inspiral, plunge-and-merger, and a spinning BH
remnant surrounded by a disk of magnetized tidal debris that
accretes onto it. During the inspiral,  the orbital separation
between the stars decreases adiabatically as the energy and
angular momentum are radiated by GWs (the radiated EM energy
during this phase is much smaller than the GW emission). Once
the quasi-circular inspiral orbit becomes unstable, the stars
plunge and merge. Depending on the total mass of the system, the
merged stars will promptly collapse or will form a HMNS. During the
last two phases of the evolution, magnetically driven outflows
and/or strong EM signals can be produced that may explain or give
new insight regarding sGRB phenomenology.

In the following section, we briefly  summarize the dynamics of the
delayed collapse cases previously performed in~\cite{Ruiz:2016rai}.
We then  describe the dynamics of the prompt collapse cases displayed
in Table~\ref{table:ID_allcases} and highlight the principle
differences with respect to the delayed case. Since the dynamics, GWs,
and EM signals are qualitative the same in the six prompt collapse cases,
we mainly discuss the merger and the final outcome of the P-Prompt-3 case.
Key results from our models are displayed in Table
\ref{table:results_allcases}.

\subsection{Delayed collapse}
\label{subsec:delayed}
As the magnetic-to-gas-pressure ratio in the NS interior is
initially small ($\beta^{-1}\ll 1$), the late inspiral phase
of the NSNS systems proceed basically unperturbed by the B-field.
The frozen-in B-field is simply dragged by the fluid stars and
the  magnetic energy $\mathcal{M}$ does not change significantly
(see Fig.~\ref{fig:NSNS_EM}). Note that recently 
  an enhancement of the magnetic energy during the early inspiral was
  reported in~\cite{Ciolfi:2017uak,Kiuchi:2014hja}. This behavior may be
  related to tidal deformation during this epoch.
  
GW emission drives the system to
the plunge--and-merger phase, and after $t-t_{\rm B}\sim 230M
\approx 3.60(M_{\rm NS}/1.8M_\odot)$ms following the B-field
insertion, the NSs come into contact and form a differentially
rotating HMNS. During merger leading delayed collapse, the magnetic
energy is steeply amplified. We find that by $t-t_{\rm merger}\sim 256M\approx
3.8(M_{\rm NS}/1.8M_\odot)$ms following the merger, $\mathcal{M}$
is amplified $\sim 12$ times its initial value (see~Fig.
\ref{fig:NSNS_EM}). Note that a similar behavior  was reported
in very high resolution simulations~\cite{Kiuchi:2014hja}, which
was attributed to both the KHI and MRI.

Once the system settles down to a quasiequilibrium HMNS, a strong
toroidal B-field is generated, mainly due winding by differential
rotation. As a result $\mathcal{M}$ is further amplified (by a
factor of $\sim 1.7$ in the P-Delayed case, and  $\sim 1.3$ in
the I-Delayed case). This behavior was  expected since the initial
B-field was chosen such that the magnetic energy may reach
equipartition with the kinetic energy during merger and HMNS
evolution, as was suggested previously in~\cite{Kiuchi:2014hja}. We
find that the wavelength~$\lambda_{\rm MRI}$ of the fastest growing
MRI is resolved by $\gtrsim 10$ grid points and it fits within the
star~\cite{Gold:2013zma}. We also find that the MRI timescale is
$\tau_{MRI}\sim \Omega^{-1}\sim$ $40-100 (M_{NS}/1.8M_\odot)^{1/2}$km
$\sim 0.13-0.33 (M_{NS}/1.8M_\odot)^{1/2}$ms (for details see~\cite{dlsss06b}).
Here $\Omega$ is the angular velocity of the HMNS.
Thus, it is likely that the MRI is properly
captured and operating in the system. Magnetic winding drives the
HMNS toward uniform rotation~\cite{Shapiro:2000zh} and, since the
rest mass of the star remnant exceeds the maximum value allowed by
uniform rotation (i.e. the ``supramassive'' limit, $M_0\approx 2.36
M_\odot(k/262.7\rm km^2)$ for $\Gamma=2$ EOS~\cite{LBS2003ApJ}),
it eventually collapses to a BH,  with mass $M_{\rm BH}\approx 2.81
M_\odot(M_{\rm NS}/1.8
M_\odot)$ and spin parameter $a/M_{\rm BH}\simeq 0.74$, surrounded
by a highly magnetized accretion disk~(see top panels of Fig.
\ref{fig:Tor_polBfield}). Just after collapse, we find that the rms
value of the B-field in the disk is $\sim 10^{15.9}(1.8M_\odot/
M_{NS})$G.

During the collapse, the inner layers of the HMNS, which contain
most of the magnetic energy, are quickly swallowed by the BH, and
thus the magnetic energy $\mathcal{M}$ steeply decreases during
$t-t_{\rm BH}\sim 63{\rm M} \approx 1(M_{\rm NS}/1.8M_\odot)$ms,
until the accretion disk settles down, after which $\mathcal{M}$
slightly decreases as the magnetized material is accreted (see
Fig~\ref{fig:NSNS_EM}). Near to the end of the simulation, the
magnetic energy is $\mathcal{M}\sim 7.2\times 10^{49}(M_{\rm NS}/
1.8M_\odot)$ergs. Similar values were  reported in very high resolution
simulations~\cite{Kiuchi:2014hja}.

\begin{figure*}
  \centering
  \includegraphics[width=1.025\textwidth,trim=0 320 0 0,clip=true]{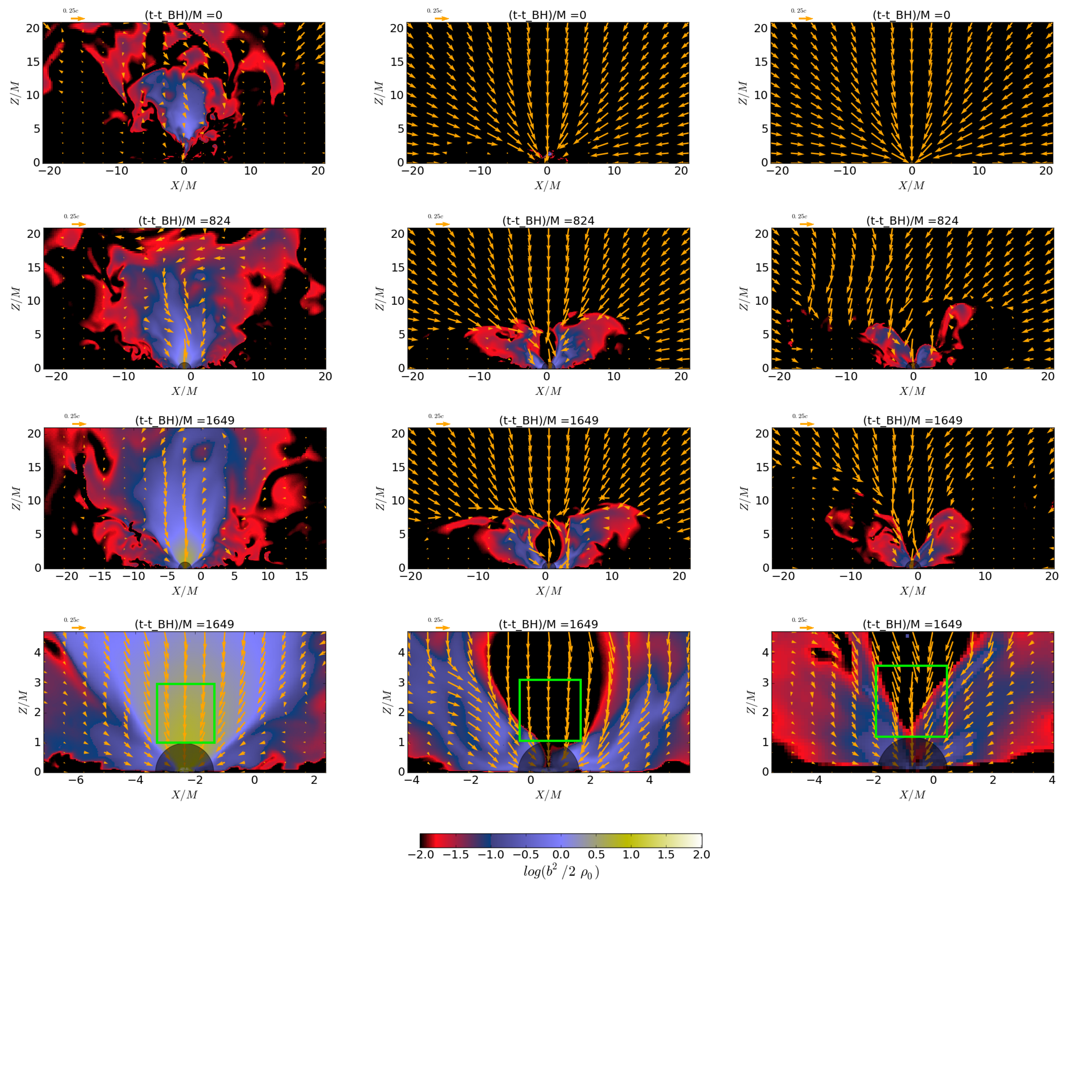}
  \caption{ \label{fig:b2rho0onxz}
    Force-free parameter $b^2/(2\,\rho_0)$ on a meridional plane at selected times for the
    P-Delayed case (left panel) and the P-Prompt-3 and I-Prompt cases (middle and right panels,
    respectively). The arrows indicate the plasma velocities, while the black semicircles
    denotes the  BH apparent horizons. Magnetically-dominated force-free areas correspond to regions
    where $b^2/(2\rho_0) \geq 1$.  The bottom row is a zoomed-in view of regions around
    the BH in the  third row. The square above the BH pole denotes the cubical region used to
    compute the average values of the force-free parameter in Fig~\ref{fig:b2overtworho0vst}.}
\end{figure*}

As the  accretion proceeds, the force-free parameter
$B^2/(8\,\pi\,\rho_0)=b^2/(2\rho_0)$  gradually grows as the regions
above the BH poles are getting cleaned out of fall-back material
(see right panel of Fig.~\ref{fig:b2rho0onxz}). Once the exterior is
magnetically-dominated, the magnetic pressure above the BH poles is
high enough to overcome the ram pressure produced by the fall-back
material. We observe that when the force-free parameter
reaches vales $\gtrsim 10$, fluid velocities begin to turn. By $t-
t_{\rm BH}\sim 2900{\rm M}\approx 45.42 (M_{\rm NS}/1.8M_\odot)$ms,
a magnetically driven outflow extends to heights  $\geq  100{\rm M}
\approx 470 (M_{\rm NS}/1.8 M_\odot)\rm km$, and an incipient jet has
been launched. Near the end of the simulation, the Lorentz factor
inside the funnel is $\Gamma_L \sim 1.2$ and thus the jet is only mildly
relativistic. However, we also find that, at that time,  the space-average
value of $b^2/(2\rho_0)$  in a cubical region with length side
of $2\,r_{BH}$ above the BH has grown to $\sim 10^{2.2}$ (see Fig.
\ref{fig:b2overtworho0vst}) and is thus becoming force-free. Since the
terminal Lorentz factor of a magnetically driven, axisymmetric jet
is comparable to this parameter~\cite{McKinney2005}, the jet may
be accelerated to higher Lorentz factors. Here $r_{BH}$ is the radius
of the BH horizon. Near the end of the simulation, the rms value
of the  the B-field is $\sim 10^{15.9} \,(1.8M_\odot/M_{NS})$G.
See Table~\ref{table:results_allcases} for the I-Delayed case.
Based on the accretion rate and the mass of the accretion disk, we
find that the disk will be accreted in $\Delta t\sim M_{\rm disk}/\dot{M}
\sim 97\rm ms$, which is consistent with  timescales of  short duration
sGRBs~\cite{Kann:2008zg}. The angular frequency  of the B-field  and
the outgoing EM luminosity are consistent with those expected from the
Blandford--Znajek mechanism~\citep{BZeffect}, as we discussed in
\cite{prs15,Ruiz:2016rai}.
%
%
\begin{center}
  \begin{table*}[th]
    \caption{Summary of main results.
      Here $M_{\rm BH}$ is the mass of the BH remnant in units of $M_\odot(
      M_{\rm NS}/1.8M_\odot)$, $a/M_{\rm BH}$ its spin parameter, $b^2/(2
      \rho_0)_{\rm ave}$ is the space-averaged value of the
      magnetic-to-rest-mass-density ratio over all the grid points inside
      a cubical region of length $2r_{BH}$ above the BH pole (see bottom panel
      of Fig.~\ref{fig:b2rho0onxz} and~Fig.~\ref{fig:b2overtworho0vst}), $r_{BH}$
      is the radius of the BH apparent horizon,  $B_{\rm rms}$ denotes the rms
      value of the B-field above the BH poles in units of $(1.8M_\odot/M_{NS})$G,
      $M_{\rm disk}/{M_0}$ is the ratio of the disk
      rest-mass to the initial total rest mass, $\dot{M}$ is the accretion rate
      computed via Eq. (A11) in~\cite{Farris:2009mt}, $\tau_{\rm disk}\sim
      M_{\rm disk}/\dot{M}$ is the disk lifetime in units of $(M_{\rm NS}/1.8M_\odot)$ms,
      $L_{\rm EM}$ is the Poynting luminosity driven by the incipient jet for the
      delayed collapse  time-averaged over the last $500{\rm M}\sim 7.3(M_{\rm NS}/
      1.8M_\odot)\rm ms$ of the evolution.
      \label{table:results_allcases}}
    \begin{tabular}{ccccccccc}
      \hline\hline
          {Case Model} & $M_{\rm BH}$ & $a/M_{\rm BH}$ & $b^2/(2\rho_0)_{\rm ave}$ & $B_{\rm rms}$ & $M_{\rm disk}/{M_0}$ & $\dot{M} (M_\odot/s)$ & $\tau_{\rm disk}$
          & $L_{\rm EM}$ $\rm erg\,s^{-1}$\\
          \hline
          P-Prompt-1  & 3.02 & 0.83  & $10^{-2}$   & $10^{14.6}$   & $0.13\%$   &  0.34  & $13.4$ & $-$           \\
          I-Prompt-1  & 3.00 & 0.83  & $10^{-6}$   & $10^{13.3}$   & $0.036\%$  &  0.08  & $15.8$ & $-$           \\
          P-Prompt-2  & 3.23 & 0.80  & $10^{-3}$   & $10^{14.5}$   & $0.085\%$  &  0.23  & $13.8$ & $-$           \\
          I-Prompt-2  & 3.22 & 0.80  & $10^{-6}$   & $10^{13.3}$   & $0.011\%$  &  0.02  & $20.6$ & $-$           \\
          P-Prompt-3  & 3.11 & 0.81  & $10^{-1}$   & $10^{15.1}$   & $0.20\%$   &  0.36  & $20.1$ & $-$           \\
          I-Prompt-3  & 3.11 & 0.81  & $10^{-1}$   & $10^{14.7}$   & $0.20\%$   &  0.37  & $19.6$ & $-$           \\
          P-Delayed   & 2.81 & 0.74  & $10^{2.2}$  & $10^{15.9}$   & $1.0\%$    &  0.33  & $97.0$ & $10^{51.3}$   \\
          I-Delayed   & 2.81 & 0.74  & $10^{1.5}$  & $10^{15.7}$   & $1.5\%$    &  0.77  & $62.3$ & $10^{50.7}$   \\
          \hline\hline
    \end{tabular}
  \end{table*}
\end{center}
%
%
\subsection{Prompt collapse}
\label{subsec:prompt}
As in the above scenario, during the NSNS inspiral the magnetic energy
$\mathcal{M}$ hardly changes, as the  dynamically unimportant, frozen-in B-field is
advected with the fluid (see Fig.~\ref{fig:NSNS_EM}). GW emission drives
the system to the plunge--and-merger phase, and after $t-t_B\sim 170M
\approx 2.78 (M_{\rm NS}/1.8M_\odot)$ms following the B-field insertion
the stars merge, forming a double core structure. The core collapses promptly to a
highly spinning  BH after $t-t_{\rm merger }\sim 80M\approx 1.3 (M_{\rm NS}/
1.8M_\odot)$ms following the merger of the two cores (see
middle panel in Fig.~\ref{fig:NSNS_ID}). The BH settles to a mass of
$M_{\rm BH}\approx 3.1M_\odot(M_{\rm NS}/1.8M_\odot)$ and spin parameter
$a/M\simeq 0.82$, consistent with~\cite{STU1,Liu:2008xy}. See Table
\ref{table:results_allcases} for the other cases.

During the NSNS merger, but before the merger of the two dense cores,
the magnetic energy $\mathcal{M}$ is quickly amplified until BH
formation~(see Fig.~\ref{fig:NSNS_EM}). In contrast to the delayed
collapse case, where the HMNS stage allows the magnetic energy to grow
a factor of $\sim 12$ (see above), we find that $\mathcal{M}$ is
amplified only~$\lesssim 4$ times its initial value.

As the low-density layers of the merging NSs wrap around the BH to
form the accretion disk, the B-field is stretched and wound, producing
a strong toroidal B-field component~(see middle and bottom panels in
Fig~\ref{fig:Tor_polBfield}). Just after collapse, the rms value of the
toroidal component of the B-field is $\sim 10^{15.1}\,(1.8M_\odot/
M_{NS})$G for the P-Prompt-3 case, and $\sim 10^{14.7}\,(1.8M_\odot
/M_{NS})$G for the I-Prompt-3 case. These values  change only slightly
during the subsequent evolution (see below). We find that the wavelength
of $\lambda_{MRI}$ is resolved by more than $\gtrsim 10$ grid points but
only partially fits in the  accretion disk~\cite{Gold:2013zma}.
As shown in the right panels of~Fig.~\ref{fig:Tor_polBfield}, the accretion disk
is not only lighter than the disk in  delayed collapse (see
Table~\ref{table:results_allcases}), but also two times smaller. It is
more difficult to properly resolve the MRI in the prompt collapse case.
Here $\Omega$ is the angular velocity of the accretion disk.
However, we do observe indications of turbulence in the disk on meridional
slices which may be produced by the existence of unstable global modes~\cite{Gammie94}.
We also find that the effective Shakura--Sunyaev $\alpha$-parameter,
as defended in~\cite{Gold:2013zma}, which associated with the magnetic
stresses is $\alpha= 0.07-0.4$, consistent with GRMHD simulations
of accretion disks around highly spinning BHs~\cite{Krolik2007}. Finally, 
we find that the MRI timescale is $\tau_{MRI}\sim $
$ 25-100\,(M_{NS}/1.8M_\odot)^{1/2}$km $\sim 0.08-0.33 (M_{NS}/1.8M_\odot)^{1/2}$ms.
Thus, it is likely that magnetically-driven turbulence is operating,
at least partially, in our system.

We follow the evolution of the BH-disk remnant for around $t-t_{\rm merger}
\sim 1650M\approx 26.74(M_{\rm NS}/1.8M_\odot)$ms after merger, which
corresponds to  $\sim 5.3$ Alfv\'en time scales. No further enhancement in the
magnetic energy was observed. In fact, as it is shown in Fig.~\ref{fig:NSNS_EM},
as the accretion proceeds, $\mathcal{M}$ gradually decreases.  Near the end
of the simulation, the magnetic energy is $\mathcal{M}\approx 2.65\times
10^{48}(M_{\rm NS}/1.8M_\odot)$ ergs, a factor of $\sim 30$ smaller than in
the delayed case.

\begin{figure}
  \centering
  \includegraphics[width=0.50\textwidth]{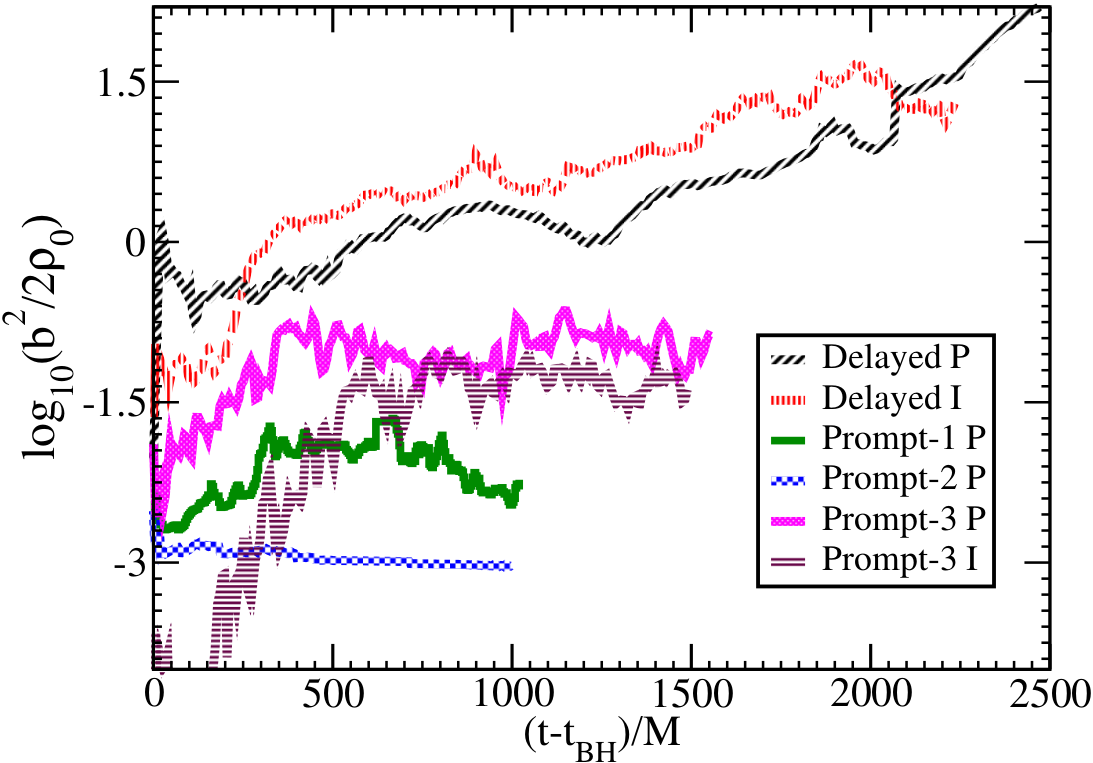}
  \caption{Average value of the force-free parameter $b^2/(2\,\rho_0)$ vs time (log scale).
    This average is computed using all the grid points
    contained in a cube of edge $2\,r_{BH}$ above the BH as shown in the bottom
    panel of Fig.~\ref{fig:b2rho0onxz}. Here  $r_{BH}$ denotes the radius of the BH.
    \label{fig:b2overtworho0vst}}
\end{figure}

As the matter above the BH poles is accreted,  meridional slices (see Fig.
\ref{fig:b2rho0onxz}) show  the expansion of regions where the force-free
parameter reaches values of $b^2/(2\rho_0)\approx 10^{-0.6}$ (quasi-magnetic
dominated regions). We observe that after $t-t_{BH}\sim 1000M\approx
16.2(M_{\rm NS}/1.8M_\odot)$ms, the magnetic pressure is high enough to push
material upward until it balances the ram pressure at a height of $\sim 8r_{BH}$
(see middle and right panels of Fig.~\ref{fig:b2rho0onxz}). After that point,
the system settles down. The quasi-magnetically dominated regions bounce up and
down above the BH poles, but they never escape. Fig.~\ref{fig:b2rho0onxz} shows
a side by side comparison of these regions in the P-Delayed case (left panel)
and P-Prompt-3 and I-Prompt-3 cases (middle and right panels, respectively) on
a meridional slice at three different times: at BH formation, at about halfway
and near the end of the simulations of the Prompt-3 cases.  As it can be seen,
at the BH formation time, there are no magnetically dominated force-free regions in
any of the two Prompt collapse cases. In contrast, in the delayed case,
these regions extend to~$\sim 10\,r_{BH}$ above the BH poles. As the
evolution proceeds, $b^2/(2\rho_0)$  gets larger in all the three cases. However,
while in the delayed collapse case these regions continuously expand
as the accretion proceeds, in the prompt  collapse cases the force-free
parameter settles down. To verify this, we compute the space-averaged value of
$b^2/(2\rho_0)$ on a cubical region of a length side $2\,r_{BH}$ just above the
BH poles during  the  evolution (see bottom panels of Fig.\ref{fig:b2rho0onxz}).
We find that the average value the force-free parameter in both the P-Prompt-3
and the I-prompt-3 cases reaches a value of $b^2/(2\rho_0)_{\rm ave}\sim 10^{-1}$
after $\sim 800M\approx 13(M_{\rm NS}/1.8M_\odot)$ms and then settles down (see
Fig.~\ref{fig:b2overtworho0vst}).
In the delayed collapse case, in contrast,  the averaged value monotonically
increases with time, whereby by $t-t_{BH}\sim 2900M \sim 42.2(M_{\rm NS}/
1.8M_\odot)$, near the end of the simulation, the force-free parameter reaches
a value of $b^2/(2\rho_0)_{\rm ave}\sim 10^{2.2}$. See Table
\ref{table:results_allcases} for the other cases.

We also compare in Fig.~\ref{fig:Tor_polBfield} the strength of the poloidal
and toroidal B-field components as well as the B-field configurations toward
the end of the P-Prompt-3 cases. In the three cases, the toroidal component
is, as expected, the dominant component in the accretion disk, while in regions
above the BH poles the poloidal component dominates. Notice that only in the
delayed collapse case does the B-field  reach  equipartition-strength
values ($\gtrsim 10^{15.7}\,(1.8M_\odot/M_{NS})$G) in both the disk and in the
funnel, which reinforces the fact that if the B-field in NSNS mergers can be
amplified to equipartition levels then, the system provides a viable model
for sGRBs~\cite{Ruiz:2016rai,Kiuchi:2014hja}. See~Table~\ref{table:results_allcases}
for the other cases.  Finally, notice that, as shown in the right panel of Fig.
\ref{fig:Tor_polBfield}, by $t-t_{BH}\sim 1650M\approx 26(M_{\rm NS}/1.8M_\odot)$ms
the winding of the B-field above the BH poles is well underway only in the P-Delayed
case. There is no evidence of that effect in any of the prompt collapse cases. So, our
results indicate that NSNS mergers can be the progenitors that power sGRBs only if
the magnetic energy can be efficiently amplified to equipartition levels
\cite{Ruiz:2016rai,Kiuchi:2014hja}, which seems to be  possible only if a transient
HMNS forms, i.e.  only in NSNS systems that lead to delayed collapse to BH.
%
%
%
\section{Conclusions}
\label{sec:conclusion}
Mergers of NSNSs have been suggested as one of the possible  progenitors
of  sGRBs~\cite{EiLiPiSc,NaPaPi,MoHeIsMa}. This hypothesis
has been reinforced by the first detection of a kilonova associate with the
system~``GRB 130603B''~\cite{Tanvir:2013pia,Berger:2013wna}. Using numerical
simulations, we have recently shown~\cite{Ruiz:2016rai} that NSNS systems
that undergo {\it delayed} collapse can launch a magnetically-sustained mildly
relativistic outflow-- an incipient jet.  The accretion time scale of the
disk and outgoing electromagnetic signals are consistent with sGRBs
as well as with the Blandford--Znajek mechanism for launching these
jets and associated Poynting luminosities~\citep{BZeffect}.

In this paper, we have performed  magnetohydrodynamic simulations in full
general relativity of different NSNS configurations that  undergo {\it prompt}
collapse (see Table~\ref{table:ID_allcases}). The stars possess a B-field that
extends from the NS interior into the exterior in some cases, or a B-field
that is confined to the NS interior. Our results show that the absence of a
HMNS epoch for prompt collapse prevents the magnetic energy from approaching
force-free values above the BH poles. This limitation inhibits the
launching of a jet.  After $t-t_{BH}\sim 1000M \sim 16.2(M_{\rm NS}/
1.8M_\odot)$ms following the collapse, we did not find any evidence of an
outflow or B-field collimation as we did in delayed collapse. At the end
of the simulations the rms value of the B-field is $\lesssim 10^{15.1}\,
(1.8M_\odot/M_{NS})$G and $b^2/(2\rho_0)\lesssim 0.1$. Our results seem to
reinforce the previous NSNS studies that claim that only NSNS systems in which
the magnetic energy reaches equipartition levels can launch magnetically-supported
jets, and may be then progenitors of sGRBs~\cite{Kiuchi:2014hja}.
Notice that, although higher resolution is required to properly capture the
KHI and  the MRI, we do not expect a significant change in the outcome.
The magnetic energy amplification due to these magnetic instabilities occurs
on an Alfven timescale ($\sim5(M_{NS}/1.8M_\odot)$ms), but, in the prompt
collapse cases, the  NSNS remnant collapses on a shorter timescale
($\sim1.3(M_{NS}/1.8M_\odot)$ms)  preventing their growth. So, 
the magnetic energy in these cases cannot reach equipartition levels
required to trigger jets.

Although our study is illustrative and not exhaustive, it suggests that
coincident detections of gravitational waves with sGRBs
may be possible only in  delayed collapse but not in the case of prompt
collapse. This finding can be  used to constrain the EOS if the masses
of these stars in the binary can be reliably
determined from measurements of the  gravitational signals  during the
pre-merger inspiral phase of a merging NSNS~\cite{Cutler:1994ys,
  Sathyaprakash:2009xs}. For example,  if the well-known binary pulsar
PSR 1913+16 merges and a GW signal is detected in coincidence with an sGRB
then we will know that the EOS  that models these
stars must have  a threshold value for prompt collapse larger than the
total mass of this binary ($\gtrsim 2.83M_\odot$). Thus, this coincident
detection  will automatically rule out the SLy and FPS EOSs,
whose threshold masses are $\sim 2.7M_\odot$ and $\sim 2.5M_\odot$,
respectively~\cite{STU2,Bauswein:2017aur}. Additionally,
measurement of the time delay between the gravitational peak and sGRBs
may provide an estimate of the initial neutron star
B-field strength~\cite{BaShSh}.


\acknowledgements
We thank Charles Gammie, Roman Gold and Vasileios Paschalidis
for useful discussions, and the Illinois Relativity group REU
team (Eric Connelly,Cunwei Fan, Patchara Wongsutthikoson and
John Simone) for assistance in creating Fig.~\ref{fig:NSNS_ID}.
This work has been supported in part by National Science Foundation
(NSF) Grants PHY-1602536 and PHY-1662211, and NASA Grants NNX13AH44G
and 80NSSC17K0070 at the University of Illinois at Urbana-Champaign.
This work made use of the Extreme Science and Engineering Discovery
Environment (XSEDE), which is supported by National Science Foundation
grant number TG-MCA99S008. This research is part of
the Blue Waters sustained-petascale computing project,
which is supported by the National Science Foundation
(awards OCI-0725070 and ACI-1238993) and the State of
Illinois. Blue Waters is a joint effort of the University
of Illinois at Urbana-Champaign and its National Center
for Supercomputing Applications.

\bibliographystyle{apsrev4-1}        
\bibliography{references}            
\end{document}